\documentclass[iop,numberedappendix,twocolappendix,twocolumn]{emulateapj} 

\usepackage{latexsym}   %Simbolos de latex
\usepackage{amsfonts}   %Fuentes de la AMS
\usepackage{amstext}    %Texto de la AMS
\usepackage{amsmath,amssymb}
\usepackage{mathtools}
\usepackage{bm}
\usepackage{MnSymbol} 
\usepackage{mathrsfs}
\usepackage{hyperref}
\hypersetup{
  colorlinks   = true, %Colours links instead of ugly boxes
  urlcolor     = blue, %Colour for external hyperlinks
  linkcolor    = blue, %Colour of internal links
  citecolor   = blue %Colour of citations
}

\begin{document}
 
\title{\bf Formal solutions for polarized radiative transfer\\ I. the DELO family}
\author{Gioele Janett\altaffilmark{1,2}, Edgar S. Carlin\altaffilmark{1}, Oskar Steiner\altaffilmark{1,3}, Luca Belluzzi\altaffilmark{1,3}}%Siddhartha Mishra\altaffilmark{2}
\email{gioele.janett@irsol.ch}%steiner@kis.uni-freiburg.de, belluzzi@irsol.ch

\affil{$^1$ Istituto Ricerche Solari Locarno (IRSOL), 6605 Locarno-Monti, Switzerland\\
$^2$ Seminar for Applied Mathematics (SAM), ETH Zurich, 8093 Zurich, Switzerland\\
$^3$ Kiepenheuer-Institut f\"ur Sonnenphysik (KIS), D-79104 Freiburg i.~Br., Germany}

\begin{abstract}
The discussion regarding the numerical integration of the polarized radiative transfer equation is still open and the comparison between the different numerical schemes proposed by different authors in the past is not fully clear. Aiming at facilitating the comprehension of the advantages and drawbacks of the different formal solvers, this work presents a reference paradigm for their characterization based on the concepts of \emph{order of accuracy}, \emph{stability}, and \emph{computational cost}. Special attention is paid to understand the numerical methods belonging to the Diagonal Element Lambda Operator family, in an attempt to highlight their specificities.
\end{abstract}
%
% %%%%%%%%%%%%%%%%%%%%%%%%%%%%%%%%%%%%%%%%
\section{Introduction}\label{sec:sec1}
% %%%%%%%%%%%%%%%%%%%%%%%%%%%%%%%%%%%%%%%%
The transfer of partially polarized light is described by a system of coupled first-order, inhomogeneous ordinary differential equations. Explicitly,
\begin{equation}
  \frac{\rm d}{{\rm d} s}\mathbf I(s)  = -\mathbf K(s)\mathbf I(s) + \boldsymbol{\epsilon}(s)\coloneqq \mathbf F(s,\mathbf I(s))\,,
\label{eq:RTE}
\end{equation}
where $s$ is the spatial coordinate measured along the ray under consideration, $\mathbf I=(I_1,I_2,I_3,I_4)^{T}\equiv(I,Q,U,V)^{T}$ is the Stokes vector,
\begin{equation*}
  \mathbf K = \begin{pmatrix}
      \eta_I &  \eta_Q &  \eta_U & \eta_V  \\
      \eta_Q &  \eta_I &  \rho_V & -\rho_U \\
      \eta_U & -\rho_V &  \eta_I & \rho_Q  \\
      \eta_V &  \rho_U & -\rho_Q & \eta_I 
               \end{pmatrix}
\label{matrix_K}
\end{equation*}
is the propagation matrix, and $\boldsymbol{\epsilon}=(\epsilon_I,\epsilon_Q,\epsilon_U,\epsilon_V)^{T}$ is the emission vector, which represents the source term. The different coefficients appearing in the propagation matrix and in the emission vector depend on the considered frequency, on the propagation direction, and on different atmospheric parameters \citep{landi_deglinnocenti+landi_deglinnocenti1985,landi_deglinnocenti1987}. For notational simplicity, the frequency dependence of the quantities is not explicitly indicated.

Analytical solutions of Equation~\eqref{eq:RTE} are available for a few simple model atmospheres only \citep{lopez1999b}, which explains the necessity of numerical schemes able to solve it. The definition of \emph{formal solution} was first introduced for the scalar problem: it is the evaluation of the radiation intensity, given knowledge of the boundary conditions and the spatial, angular, and frequency dependence of the opacity and the emissivity at a discrete set of points \citep{mihalas1978,auer2003}. The generalization to the polarized case consists in substituting radiation intensity, opacity, and emissivity by Stokes vector, propagation matrix, and emission vector, respectively.

The formal solution of the radiative transfer equation is a key step of iterative schemes for solving the nonlinear full radiative transfer problem, where the atomic system and the radiation field interact in non-local thermodynamic equilibrium (non-LTE) conditions. From the computational point of view, the formal solution is, in most cases, the slowest part of the iterative scheme \citep[e.g.,][]{stepan+trujillo_bueno2013}. Moreover, large magnetohydrodynamic simulations of stellar atmospheres call for massive synthesis of Stokes profiles. Therefore, the requirement for the numerical method is to be as accurate and as fast as possible.  The effort of the community has produced an extensive literature on the different formal solvers, the major contributions being summarized in Table \ref{tab:history}. The quest of the ``best formal solver'' available is still open and the comparison between the different numerical schemes provided by the community is somehow confusing. 

This paper aims to give a structured overview over formal solvers, in particular over those belonging to the Diagonal Element Lambda Operator (DELO) family, and to clarify some incoherences found in the literature. Section~\ref{sec:sec2} presents a reference paradigm for the characterization of formal solvers. The concepts of \emph{order of accuracy}, \emph{stability} and \emph{computational cost} are briefly presented and used to characterize a reference numerical scheme, namely the trapezoidal method. Section~\ref{sec:sec3} provides an introduction to exponential integrators, a class of numerical methods for the solution of differential equations. A simple description of this class is given, in an attempt to highlight its specific features and its paternity of the DELO methods. In Section~\ref{sec:sec4}, the well known DELO family is presented and characterized. Some new methods belonging to this family are introduced and compared to the already existing ones. Finally, Section~\ref{sec:sec5} provides remarks and conclusions, with a view on future work.
\begin{table*}
\caption{List of formal solvers proposed by different authors}
\centering
\begin{tabular}{|l|l|l|}
\hline
\emph{Year} & \emph{Method} & \emph{Proposed by}\\%\hline%\midrule
\hline
1974 & Runge-Kutta-Merson  & \citet*{wittmann1974}\\ 
1976 & Runge-Kutta 4  &  \citet*{landi_deglinnocenti1976}\\ 
1985 & Piecemeal Evolution Operator & \citet*{landi_deglinnocenti+landi_deglinnocenti1985}\\ 
1989 & Zeeman Feautrier & \citet* {rees+al1989}\\ 
1989 & DELO-linear & \citet*{rees+al1989}\\
1998 & (cubic) Hermitian & \citet*{bellot_rubio+al1998}\\
1999 & DIAGONAL & \citet*{lopez1999c}\\
2003 & DELOPAR & \citet*{trujillo_bueno2003}\\
2013 & (quadratic and cubic) DELO-B{\'e}zier & \citet*{delacruz_rodriguez+piskunov2013}\\
2013 & BESSER & \citet*{stepan+trujillo_bueno2013}\\
2016 & Piecewise Continuous & \citet*{steiner2016}\\
\hline
%\bottomrule
\end{tabular}
\label{tab:history}
\end{table*}
%
% %%%%%%%%%%%%%%%%%%%%%%%%%%%%%%%%%%%%%%%%
\section{Characterization of formal solvers}\label{sec:sec2}
% %%%%%%%%%%%%%%%%%%%%%%%%%%%%%%%%%%%%%%%%
In order to be able to answer the question ``which is the best formal solver?'', one first has to fix some criteria for judging the different numerical schemes. Briefly discussing a method, claiming some strong points, and showing them with a few specific examples, is not the best way to proceed. The literature about the numerical approach to ordinary differential equations is particularly broad and great efforts have been directed toward the characterization of the different methods. This section aims to give an overview on this characterization, in order to facilitate the comprehension of the advantages, weaknesses, and possible incoherences of the already existing formal solvers and of those yet to come. To ease the appreciation, a very common and simple method is presented and analyzed, namely the trapezoidal method.
\subsection{Exempli gratia: the trapezoidal method}\label{subsec:1.1}
A spatial grid $\{s_k\}$ $(k=0,\dots,N)$ is introduced, discretizing the ray path. The spatial coordinate $s$ and the index $k$ increase along the propagation direction. For a given grid point $s_k$, the points $s_{k-1}$ and $s_{k+1}$ represent the \emph{upwind} and \emph{downwind} points, respectively, guaranteeing $s_{k-1} \le s_k \le s_{k+1}$. Applying the fundamental theorem of calculus to Equation~\eqref{eq:RTE} in the interval $\left[s_k,s_{k+1}\right]$, one obtains
\begin{equation}
\mathbf I(s_{k+1})-\mathbf I(s_k)=\int_{s_k}^{s_{k+1}}\mathbf F(s,\mathbf I(s)){\rm d}s\,.
\label{fundamental_theorem_calculus}
\end{equation}
The integral can be approximated by different numerical quadratures and one possibility is to use the trapezoidal rule, i.e.,
\begin{equation*}
\int_{a}^{b}f(x){\rm d}x\approx\frac{b-a}{2}\left[f(a)+f(b)\right]\,.
\end{equation*}
Approximating the integral in Equation~\eqref{fundamental_theorem_calculus} in terms of the trapezoidal rule, one recovers
{\small
\begin{equation}
\mathbf I(s_{k+1})-\mathbf I(s_k)\approx\frac{\Delta s_k}{2}\left[\mathbf F(s_k,\mathbf I(s_k))+\mathbf F(s_{k+1},\mathbf I(s_{k+1}))\right]\,,
\label{trapezoidal_rule}
\end{equation}}\noindent
where $\Delta s_k=s_{k+1}-s_k$ is the cell width. The numerical approximation of a certain quantity at node $s_k$ is indicated by substitution of the explicit dependence on $s$ with the subscript $k$, for instance,
\begin{equation*}
\mathbf I_k \approx \mathbf I(s_k)\,.
\end{equation*}
Inserting numerical quantities for $\mathbf I$, $\mathbf K$, and $\boldsymbol{\epsilon}$ in Equation~\eqref{trapezoidal_rule} and applying some algebra, one recovers the implicit linear system
\begin{equation}
\mathbf{\Phi}_{k+1}\mathbf I_{k+1}=\mathbf{\Phi}_k\mathbf I_k+\mathbf{\Psi}_{k+1}+\mathbf{\Psi}_k\,,
\label{trapezoidal}
\end{equation}
where the matrices $\mathbf{\Phi}_k$ and $\mathbf{\Phi}_{k+1}$ and the vectors $\mathbf{\Psi}_k$ and $\mathbf{\Psi}_{k+1}$ are given by
\begin{align*}
\mathbf{\Phi}_k&=\mathbf{1}-\frac{\Delta s_k}{2}\mathbf K_k\,,\\
\mathbf{\Phi}_{k+1}&=\mathbf{1}+\frac{\Delta s_k}{2}\mathbf K_{k+1}\,,\\
\mathbf{\Psi}_k&=\frac{\Delta s_k}{2}\boldsymbol{\epsilon}_k\,,\\
\mathbf{\Psi}_{k+1}&=\frac{\Delta s_k}{2}\boldsymbol{\epsilon}_{k+1}\,.
\end{align*}
Given the upwind Stokes vector $\mathbf I_k$ at node $s_k$, one solves the implicit linear system given by Equation~\eqref{trapezoidal}, finding the emergent Stokes vector $\mathbf I_{k+1}$ at node $s_{k+1}$.

The trapezoidal method is therefore classified as an implicit method and belongs to both the famous classes of the Runge-Kutta methods and the linear multistep methods.
\subsection{Order of accuracy}\label{subsec:1.2}
In order to recover a good approximation, one tries to maintain as small an error as possible. When discussing numerical schemes, one usually refers to two different kinds of errors: the \emph{local truncation error} and the \emph{global error}.

The \emph{local truncation error} is the error introduced by the numerical scheme in a single step, assuming the exact solution at the precedent step. Considering a scalar initial value problem (IVP) of the general form
\begin{equation*}
\begin{aligned}
y'(t)&=f(t,y(t))\,,\\
y(t_0)&=y_0\,,
\end{aligned}
\end{equation*}
supplied by the discrete grid $\{t_k\}$ $(k=0,\dots,N)$, the local truncation error is defined by
\begin{equation*}
L_k=\Vert y_k-y(t_k)\Vert\,,\text{ with }y_{k-1}=y(t_{k-1})\,,
\end{equation*}
where $y(t_{k-1})$ and $y(t_k)$ are the exact solutions at nodes $t_{k-1}$ and $t_k$, respectively, and $y_k$ represents the numerical approximation at node $t_k$ after performing the single step from $t_{k-1}$ to $t_k$. The operator $\Vert \cdot \Vert$ represents any suitable vector norm. A numerical method for an ordinary differential equation has order of accuracy $p$ if the local truncation error, which usually depends on the step size denoted by $\Delta t$, satisfies
\begin{equation}
L_k \approx O(\Delta t^{p+1})\,,\;\text{with }p\ge1\,.
\label{local_error}
\end{equation}
Hence, the larger the order of accuracy, the faster the error is reduced as $\Delta t$ decreases. 

By contrast, the \emph{global error} is defined as
\begin{equation*}
E_N=\Vert y_N-y(t_N)\Vert\,,
\end{equation*}
and represents the accumulation of the local truncation error over all the $N$ steps. If the same amount of error is produced at each step, i.e., without any amplification of errors over subsequent steps, then a local truncation error of $O(\Delta t^{p+1})$ implies a global error of $O(\Delta t^p)$. Explicitly
\begin{equation}
E_N \approx\sum_{k=1}^N L_k \approx C \cdot \Delta t^{p}+O(\Delta t^{p+1})\,,\;\text{with }p\ge1\,,
\label{global_error}
\end{equation}
using Equation~\eqref{local_error} and $N\propto1/\Delta t$. The constant $C$ typically depends on the exact solution and on other parameters of the numerical scheme, but is strictly independent of $\Delta t$. The concept of amplification of errors is related to the idea of stability, which will be shortly discussed. The intuitive connection between local truncation error and global error given by Equation~\eqref{global_error} can be generalized to any suitable discrete grid \citep[e.g.,][]{deuflhard2002}, including logarithmically spaced grids. In the case of non-uniform discrete grids, the step size $\Delta t$ must be replaced by the maximal step size given by
\begin{equation*}
\Delta \hat t=\max_{k=0,\dots,N-1} \Delta t_k\,,
\end{equation*}
with $\Delta t_k = t_{k+1}-t_k$. The expression ``suitable discrete grid'' indicates that the maximal step size is inversely proportional to the total number of grid points, i.e., $\Delta \hat t\propto N^{-1}$.

The power law presented in Equation~\eqref{global_error} implies a linear relationship between the logarithms of the global error $E_N$ and the step size $\Delta t$, i.e.,
\begin{equation*}
\log (E_N) \approx p \cdot \log (\Delta t)+\tilde C\,,
\end{equation*}
where $\tilde C = \log (C)$. This relation can be appreciated in a log-log plot, where the resulting straight line, often called the signature or error curve, should show a slope equal to $p$. The trapezoidal method is well known to be second-order accurate \citep[e.g.,][]{frank2008} and an explicit example, showing its signature, will be presented later in this paper.

Note, however, that definitions \eqref{local_error} and \eqref{global_error} assume sufficiently small $\Delta t$ in order to avoid the \emph{pre-asymptotic} behavior, i.e., the fact that for large step sizes the data points are more scattered and do not necessarily follow the power law. For instance, a global error of the form $E_N \approx C_1 \cdot \Delta t+C_2 \cdot \Delta t^2+ C_3 \cdot \Delta t^3$ is dominated by the first term for small enough $\Delta t$, but for large step sizes higher-order terms may tangibly contribute, depending on their constants $C_2$ and $C_3$.
\subsection{Stability}\label{subsec:1.3}
For a numerical scheme, \emph{stability} means that any numerical error introduced at some stage does not blow up in the subsequent steps of the method. The concept of stability is often related to the concept of stiffness. A differential equation is said to be stiff when some numerical methods have to take an extremely small step size to achieve convergence. Therefore, step size control is also based on stability requirements, because instabilities lead to a deterioration of accuracy. More details about the concept of stability in the numerical treatment of ordinary differential equations can be found, for instance, in \cite{hackbusch2014}.

There are different ways to determine the stability of a numerical scheme and the stability analysis often depends on the considered class of methods. In the following, the common class of the Runge-Kutta methods is considered. The stability of a numerical method is often deduced through the simple autonomous scalar IVP given by
\begin{equation}
\begin{aligned}
y'(t)&=\lambda y(t)\,,\\
y(0)&=y_0\,,
\label{IVP}
\end{aligned} 
\end{equation}
with $\lambda \in \mathbb{C}$. The solution $y(t)= y_0e^{\lambda t}$ converges to zero as $t \rightarrow \infty$ for $\operatorname{Re}(\lambda)< 0$. A Runge-Kutta method applied to the IVP \eqref{IVP} can be recast into the form
\begin{equation*}
y_{k+1}=\phi(\lambda\Delta t)y_k\,,
\end{equation*}
where $\phi$ is called the \emph{stability function}. The numerical method is said to be stable if, applied to the IVP \eqref{IVP}, it converges to zero for $k \rightarrow \infty$ and this condition is equivalent to
\begin{equation}
\Vert\phi(\lambda\Delta t)\Vert<1\,.
\label{stability_condition}
\end{equation}
Intuitively, this guarantees that any perturbation in the solution is attenuated with the recursive numerical integration. The stability of a method is therefore related to both the step size $\Delta t$ and the eigenvalue $\lambda$, more precisely to the term $\lambda\Delta t$. The \emph{stability region} of a Runge-Kutta method is defined as the set of complex values $\lambda\Delta t$ for which Equation~\eqref{stability_condition} is satisfied.

The stability analysis for the scalar problem given by Equation~\eqref{IVP} can be easily generalized to the linear system of ordinary differential equations given by
\begin{equation}
\begin{aligned}
\mathbf{y}'(t)&= \mathbf{A} \mathbf{y}(t)\,,\\
\mathbf{y}(0)&=\mathbf{y}_0\,,
\label{IVP2}
\end{aligned} 
\end{equation}
where the $d\times d$ matrix $\mathbf{A}$ has a basis of eigenvectors corresponding to the eigenvalues $\lambda^{(1)},\dots,\lambda^{(d)}$. Equation~\eqref{IVP2} formally corresponds to the homogeneous version of Equation~\eqref{eq:RTE}. %The source term is not included in the stability analysis, because it does not play any role in the amplification of errors over subsequent steps.
The emission term is deliberately omitted in the stability analysis because of its locality. In fact, the emission term does not affect the propagation of the information, preventing any contribution to the amplification of errors. As shown in \cite{frank2008}, a Runge-Kutta scheme applied to Equation~\eqref{IVP2} is stable if and only if it guarantees stability once applied to Equation~\eqref{IVP}, with $\lambda$ representing any eigenvalue of $\mathbf{A}$. It is important to mention the fact that the eigenvalues of the propagation operator $-\mathbf K$ in Equation~\eqref{eq:RTE} have always negative real parts. Therefore, A-stability (see below) is a sufficient condition to avoid any instability problem in the formal solution.

\begin{figure*}
\centering
\includegraphics[width=1.\textwidth]{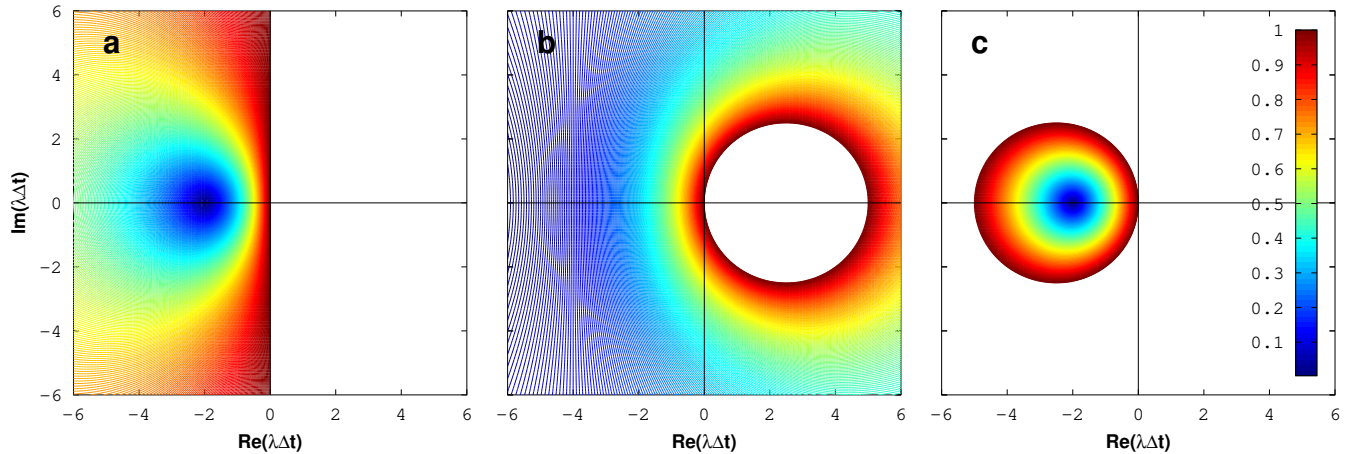}
  \caption{The stability region for the trapezoidal method for {\bf a)} $\lambda=\lambda_k=\lambda_{k+1}$, {\bf b)} $\lambda=\lambda_k=\frac{1}{5}\lambda_{k+1}$, and {\bf c)} $\lambda=\frac{1}{5}\lambda_k=\lambda_{k+1}$. Colors indicate the absolute values of the stability function $|\phi_{\text{\tiny T}}|$ given by Equation~\eqref{stability_function}. The region where the stability condition is not satisfied, i.e., $|\phi_{\text{\tiny T}}|>1$, is indicated in white.}
\label{stability_trapezoidal}
\end{figure*}

Applying the trapezoidal method to the IVP \eqref{IVP}, one recovers the following stability function
\begin{equation*}
\phi_{\text{\tiny T}}(\lambda\Delta t)=\frac{1+\lambda\Delta t/2}{1-\lambda\Delta t/2}\,.
\end{equation*}
The stability region for the trapezoidal method is then given by the condition \eqref{stability_condition} and it is usually displayed as presented in Figure~\ref{stability_trapezoidal}a. If the stability region of a numerical method contains the whole left-hand side of the complex plane, as in the case of Figure~\ref{stability_trapezoidal}a, then the numerical scheme is said to be A-stable. A broad and exhaustive literature is dedicated to the determination of the specific stability regions for the different numerical methods \citep{collatz1966,dahlquist1963}, but a deep digression would stray from the main aim of this paper.
% %
% \begin{figure}
% \centering
% \plotone{figs/stability_trapezoidal.eps}
%   \caption{The stability region for the trapezoidal method. Colors indicate the values of $|\phi_{\text{\tiny T}}|$.}
% \label{stability_trapezoidal}
% \end{figure}
% %
%

A strong limitation of this simplified stability analysis is the assumption of a constant eigenvalue $\lambda$ in Equation~\eqref{IVP}. A less restrictive analysis shows that variations of $\lambda$ along the integration path could affect the stability region of the numerical method. For example, the stability function of the trapezoidal method in the interval $\left[t_k,t_{k+1}\right]$ reads
\begin{equation}
\phi_{\text{\tiny T}}(\lambda_k,\lambda_{k+1},\Delta t_k)=\frac{1+\lambda_k\Delta t_k/2}{1-\lambda_{k+1}\Delta t_k/2}\,,
\label{stability_function}
\end{equation}
where $\lambda_k$ and $\lambda_{k+1}$ are the eigenvalues at the positions $t_k$ and $t_{k+1}$, respectively. Thus, the stability region depends on both eigenvalues $\lambda_k$ and $\lambda_{k+1}$, as illustrated in Figure~\ref{stability_trapezoidal}. However, the variation of the eigenvalue $\lambda$ along the integration path strongly depends on the spatial scale. In particular, a conversion from geometrical height to optical depth (see Appendix~\ref{appendix:C}) mitigates fluctuations of the eigenvalues of the propagation operator $-\mathbf K$, supporting the assumption of a constant eigenvalue~$\lambda$ in the stability analysis.
\subsection{Computational cost}\label{subsec:1.4}
%
%When designing or choosing a program, an important point is the amount of computational time and data storage necessary to execute it \citep[see for instance][]{goldreich2008}. The bottlenecks for programs have changed with technology. When memory became comparatively cheap, the bottleneck moved to floating-point operations, promoting algorithms designed to minimize the number of floating-point operations. Nowadays, the main bottleneck seems to be the transfer of data from memory to the processor \citep{schorghofer2015}.
When designing or choosing a numerical scheme, an important point is the amount of computational time and data storage necessary to execute it \citep[see for instance][]{goldreich2008}. A suitable parallelization strategy of the radiative transfer problem includes the use of multiple central processor unit (CPU) cores and parallelization via domain decomposition and/or in the frequency domain \citep{stepan+trujillo_bueno2013}. The \emph{computational cost} may act as an important factor in the choice of the appropriate formal solver, especially for large scale applications in which the repetitive integration of the radiative transfer equation plays a leading role.

The computational cost analysis of numerical schemes must be based on the premise that it depends on the specific coding, programming language, compiler, and computer architecture. Therefore, the interpreted Octave language used in this paper is not suitable to reliably determine time costs. Nevertheless, one can make some objective considerations. First of all, common sense suggests keeping the algorithm as sleek as possible, avoiding any unnecessary superstructure. Second, basic floating-point operations are carried out directly on the CPU, whereas elementary functions are usually emulated on a higher level. Correspondingly, the evaluation of, e.g., an exponential is 10-40 times more expensive than a floating-point multiplication \citep{schorghofer2015}. A third remark is made on the difference between explicit and implicit schemes. An explicit one-step method calculates the updated numerical value $y_{k+1}$ directly from the precedent value $y_k$, i.e.,
\begin{equation*}
y_{k+1}=f(y_k)\,,
\end{equation*}
while implicit one-step methods find $y_{k+1}$ by solving an equation of the type
\begin{equation*}
g(y_k,y_{k+1})=0\,,
\end{equation*}
which results in the additional solution of a $4\times4$ implicit linear system, when considering Equation~\eqref{eq:RTE}.

Concerning the data storage cost, a simple consideration can be made. In the short characteristic strategy the formal solver integrates step by step the radiative transfer equation along the ray path, using only local atmospheric quantities \citep{auer2003}. Therefore, the small amount of information retained avoids any data storage problem.
% 
% \begin{table}
% \caption{Execution times for different operations}
% \setlength{\tabcolsep}{5pt}\renewcommand{\arraystretch}{1.5}
% \centering
% \begin{tabular}{|l|c|c|}\hline%\toprule
% \emph{Operation}	& \emph{Cost}\\
% \hline %\midrule
% integer addition, subtraction, multiplication	& $< 1$ \\
% float addition, subtraction, multiplication	& 1 \\
% float division		& 2-6 \\
% integer division	& 4-10 \\
% square root		& 5-20 \\
% sine, cosine, tangent, logarithm, exponential	& 10-40 \\\hline %\bottomrule
% \end{tabular}
% \label{tab:cost}
% \end{table}
% %
\section{Exponential integrators}\label{sec:sec3}
% %%%%%%%%%%%%%%%%%%%%%%%%%%%%%%%%%%%%%%%%
Exponential integrators form a class of numerical methods for ordinary differential equations. This class is based on the exact integration of the linear part of the IVP, aiming at reducing the stiffness of the differential equation.

In order to present this class, one considers the following general IVP
\begin{equation}
\begin{aligned}
\mathbf{y}'(t)&= \mathbf{G}(t,\mathbf{y}(t))\,,\\
\mathbf{y}(t_0)&=\mathbf{y}_0\,,
\label{IVP_exp}
\end{aligned} 
\end{equation}
which is equivalent to Equation~\eqref{eq:RTE} given the initial value~$\mathbf I_0$. One splits $\mathbf{G}$ into linear and nonlinear contributions, i.e.,
\begin{equation*}
\mathbf{G}(t,\mathbf{y}(t)) = \mathbf{L}\mathbf{y}(t) + \mathbf{N}(t,\mathbf{y}(t))\,,
\end{equation*}
where the matrix $\mathbf{L}$ does not depend on the variable $t$ and the nonlinear term is given by $\mathbf{N}=\mathbf{G}-\mathbf{L}\mathbf{y}$. Equation~\eqref{IVP_exp} is then recast in the form 
\begin{equation}
\begin{aligned}
\left[\tfrac{\rm d}{{\rm d}t}-\mathbf{L}\right]\mathbf{y}(t)&= \mathbf{N}(t,\mathbf{y}(t))\,,\\
\mathbf{y}(t_0)&=\mathbf{y}_0\,,
\label{IVP_exp2}
\end{aligned} 
\end{equation}
and the exact integration of the linear part in the interval $\left[t_0,t\right]$ yields the \emph{variation of constants} formula
\begin{equation}
\mathbf{y}(t)=e^{\mathbf{L}(t-t_0)}\mathbf{y}_0+\int_{t_0}^{t}e^{\mathbf{L}(t-x)}\mathbf{N}(x,\mathbf{y}(x)){\rm d}x\,.
\label{variation_of_constant}
\end{equation}
The integral in Equation~\eqref{variation_of_constant} has to be numerically approximated and a large variety of different options is available, e.g., Runge-Kutta discretizations as explained by \cite{cox2002}. Moreover, for a non-diagonal matrix $\mathbf{L}$, the evaluation of the matrix exponential often requires an approximation and one has to combine the integrator with well-chosen algorithms from numerical linear algebra.

In the following section, a specific strategy is applied, where only the nonlinear term $\mathbf{N}$ is approximated and the exponential operator is treated exactly.
%
% %%%%%%%%%%%%%%%%%%%%%%%%%%%%%%%%%%%%%%%%
\section{The DELO family}\label{sec:sec4}
% %%%%%%%%%%%%%%%%%%%%%%%%%%%%%%%%%%%%%%%%
In this section, a particular family of methods belonging to the class of exponential integrators is presented. The first method of this family applied to the formal solution for polarized light was proposed by \cite{rees+al1989} under the name of DELO. Thereafter, a second version by \cite{trujillo_bueno2003} took the appellative DELOPAR. Additional improvements, in terms of B\'ezier interpolations, were recently provided by \cite{delacruz_rodriguez+piskunov2013} and by \cite{stepan+trujillo_bueno2013}.

As exhaustively explained by \cite{guderley1972}, this technique takes into account analytically the diagonal elements of the propagation matrix $\mathbf K$, aiming to remove stiffness from the problem. Therefore, the well-known radiative transfer equation for polarized light given by Equation~\eqref{eq:RTE} is brought in the form given by Equation~\eqref{IVP_exp2}, with the additional constraint of a diagonal matrix $\mathbf{L}$. This reformulation is facilitated by the fact that the diagonal elements of the propagation matrix are all identical. Replacing the coordinate $s$ by the optical depth $\tau$ defined by
\begin{equation}
{\rm d}\tau=-\eta_I(s){\rm d}s\,,
\label{opt_depth}
\end{equation}
one recasts Equation~\eqref{eq:RTE} into
\begin{equation}
\left[\tfrac{\rm d}{\rm d\tau}-\mathbf{1}\right]\mathbf I(\tau)=-\pmb{\mathscr S}(\tau,\mathbf I(\tau))\,,
\label{RTE_delo}
\end{equation}
where $\mathbf 1$ represents the $4\times 4$ identity matrix. The quantity $\pmb{\mathscr S}$ is the \emph{effective source function} and it is defined by
\begin{equation*}
\pmb{\mathscr S}(\tau,\mathbf I(\tau))=-\pmb{\mathscr K}(\tau)\mathbf I(\tau)+\tilde{\boldsymbol{\epsilon}}(\tau)\,,
\end{equation*}
with the modified propagation matrix $\pmb{\mathscr K}=\mathbf K/\eta_I-\mathbf{1}$ (whose diagonal elements are all equal to zero) and the modified emission vector $\tilde{\boldsymbol{\epsilon}}=\boldsymbol{\epsilon}/\eta_I$.

Observe that the optical depth scale is in fact defined so that the coordinate $\tau$, defined in Equation~\eqref{opt_depth}, decreases along the ray path, thus $\tau\le\tau_0$. Providing the upwind Stokes $\mathbf I_0=\mathbf I(\tau_0)$, the operator on the left of Equation~\eqref{RTE_delo} is inverted leading to the formula
\begin{equation}
\mathbf I(\tau)=e^{(\tau-\tau_0)}\mathbf I_0-\int_{\tau_0}^{\tau}e^{(\tau-x)}\pmb{\mathscr S}(x,\mathbf I(x)){\rm d}x\,,
\label{RTE_delo_solution}
\end{equation}
which is analogous to Equation~\eqref{variation_of_constant}.

As anticipated, different numerical quadratures of the integral in Equation~\eqref{RTE_delo_solution} lead to different numerical schemes. In particular, the DELO technique approximates the effective source function $\pmb{\mathscr S}$ by a polynomial $\mathbf P_q$ of degree $q$ inside the integration interval, i.e.,
\begin{equation}
\pmb{\mathscr S}(\tau,\mathbf I(\tau))\approx\mathbf P_q(\tau,\mathbf I(\tau))\,.
\label{polynomial_approx}
\end{equation}
Observing that $\tau_{k+1} \le \tau_k$ and evaluating Equation~\eqref{RTE_delo_solution} in the interval $[\tau_k,\tau_{k+1}]$, one obtains
\begin{equation}
\mathbf I_{k+1}=e^{-\Delta\tau_k}\mathbf I_k-\int_{\tau_k}^{\tau_{k+1}}e^{(\tau_{k+1}-\tau)}\mathbf P_q(\tau,\mathbf I(\tau))\rm d\tau\,,
\label{delo_numerical}
\end{equation}
where $\Delta\tau_k=\tau_k-\tau_{k+1}$. The integral can then be solved by parts, yielding an implicit or explicit linear system for the Stokes vector $\mathbf I_{k+1}$.

As explained by \cite{guderley1972}, the local truncation error is due to the fact that the effective source function is approximated by a polynomial of degree $q$. According to \cite{henrici1962}, this approximation results in the following local truncation error
\begin{equation}
L_k[q] \approx O(\Delta\tau_k^{q+2})\,,\text{ for }q\ge1\,,
\label{delo_accuracy}
\end{equation}
and, from definition \eqref{local_error}, a DELO method involving a polynomial $\mathbf P_q$ should show an order of accuracy equal to $q+1$. Equation~\eqref{delo_accuracy} is not defined for $q=0$, i.e., for a constant approximation of the effective source function. In this case, the local truncation error corresponds to the one for $q=1$, following a behavior similar to the implicit midpoint method \citep{deuflhard2002}. Moreover, the evaluation of the quantity $\Delta\tau_k$ plays a fundamental role in the accuracy of the numerical scheme, as explained in Appendix~\ref{appendix:C}.

As already mentioned, the DELO strategy is thought to remove stiffness from the problem. In fact, a method based on Equation~\eqref{RTE_delo_solution} tends to A-stability for a vanishing modified propagation matrix. Therefore, a sufficiently small matrix $\pmb{\mathscr K}$ should imply a rather wide stability limit. However, the simple stability analysis presented in the previous section cannot be applied to the present family of methods, because of the two different contributions from the exponential terms and the modified propagation matrix. Nevertheless,  some indications can be deduced if the simpler scalar case is analyzed. \cite{guderley1972} show that the DELO strategy increases stability with respect to the corresponding Adams-Bashford methods \citep{deuflhard2002}.

On the other hand, the second-order accurate trapezoidal method already guarantees A-stability, which dispenses from a stability improvement for formal solvers having an order of accuracy $p\le 2$.
\vspace{0.5cm}
\subsection{DELO-constant, DELO-linear and DELO-parabolic}\label{subsec:3.1}
In a very general way, the polynomial $\mathbf P_q$ in Equation~\eqref{polynomial_approx} is obtained by the Lagrangian interpolation
\begin{equation}
\pmb{\mathscr S}(\tau,\mathbf I(\tau))\approx\sum_{i=k-q+1}^{k+1}\pmb{\mathscr S}_i\ell_i(\tau)\,,
\label{lagrange}
\end{equation}
where $\pmb{\mathscr S}_i=-\pmb{\mathscr K}_i\mathbf I_i+\tilde{\boldsymbol{\epsilon}}_i$ and the Lagrange basis polynomials $\ell_i$ are given by
\begin{equation*}
\ell_i(\tau)=\prod_{\substack{k-q+1\le m \le k+1\\m\neq i}}\frac{\tau-\tau_{m}}{\tau_{i}-\tau_{m}}\,.
\end{equation*}
Here $k$ indicates an arbitrary node on the discretized ray path. The integral in Equation~\eqref{delo_numerical} can then be solved by parts, yielding an implicit linear system of the form
\begin{equation}
\mathbf{\Phi}_{k+1}\mathbf I_{k+1}=\sum_{i=k-q+1}^{k}\mathbf{\Phi}_i\mathbf I_i+\sum_{i=k-q+1}^{k+1}\mathbf{\Psi}_i\,,
\label{delo_recursive}
\end{equation}
where the different coefficients $\mathbf{\Phi}_i$ and $\mathbf{\Psi}_i$ depend on the chosen polynomial $\mathbf P_q$ and on the numerical values $\pmb{\mathscr K}_i$ and $\tilde{\boldsymbol{\epsilon}}_i$.
Provided the previous Stokes $\mathbf I_i$ for $i=k-q+1,\dots,k$, the linear system \eqref{delo_recursive} can be solved to obtain the numerical approximation $\mathbf I_{k+1}$ at $\tau_{k+1}$. Therefore, once given the boundary condition $\mathbf I_0$ at $\tau_0$, the recursive application of Equation~\eqref{delo_recursive} provides the emergent Stokes vector $\mathbf I_N$ at the end of the ray path.

\begin{figure*}
\centering
\includegraphics[width=1.\textwidth]{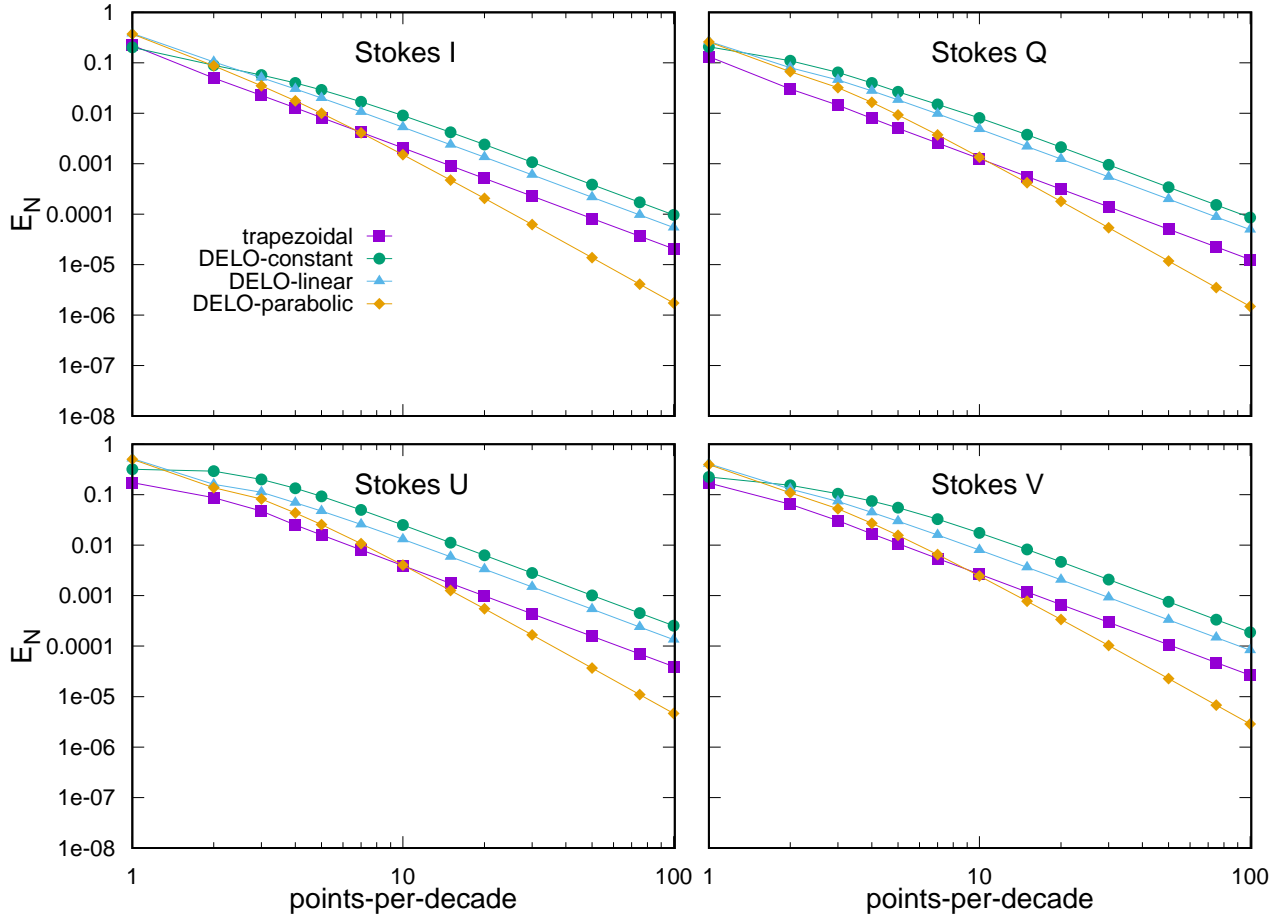}
  \caption{The log-log representation of the global error for the Stokes vector components $I,Q,U$, and $V$ as a function of the number of points-per-decade of the continuum optical depth for the trapezoidal, DELO-constant, DELO-linear, and DELO-parabolic methods. The considered atmospheric model is described in Appendix~\ref{appendix:D} and the global error is computed as shown in Appendix~\ref{appendix:A}.}
\label{fig:delo1}
\end{figure*}

As first choice, the effective source function is assumed constant inside the interval $[\tau_k,\tau_{k+1}]$, i.e.,
\begin{equation*}
\pmb{\mathscr S}(\tau,\mathbf I(\tau))\approx\pmb{\mathscr S}_{k+\frac{1}{2}}\,,
\end{equation*}
and can be approximated by the midpoint rule
\begin{equation*}
\pmb{\mathscr S}_{k+\frac{1}{2}}\approx\frac{\pmb{\mathscr S}_k+\pmb{\mathscr S}_{k+1}}{2}\,.
\label{delo_midpoint}
\end{equation*}
Replacing the polynomial $\mathbf P_q$ in Equation~\eqref{delo_numerical} by the constant approximation described above, one can calculate the integral and after some algebra obtain an implicit linear system formally identical to Equation~\eqref{trapezoidal}, i.e.,
\begin{equation}
\mathbf{\Phi}_{k+1}\mathbf I_{k+1}=\mathbf{\Phi}_k\mathbf I_k+\mathbf{\Psi}_{k+1}+\mathbf{\Psi}_k\,.
\label{delo_constant}
\end{equation}
The method described by Equation~\eqref{delo_constant}, which might be called DELO-constant, is second-order accurate, as shown in Figure \ref{fig:delo1}. The explicit values of the coefficients $\mathbf{\Phi}_k$, $\mathbf{\Phi}_{k+1}$, $\mathbf{\Psi}_k$, and $\mathbf{\Psi}_{k+1}$ are provided in Appendix~\ref{appendix:B}.

The next step is to obtain a relation between $\mathbf I_k$ and $\mathbf I_{k+1}$ when approximating the effective source function by a linear interpolation, i.e., Equation~\eqref{lagrange} with $q=1$,
\begin{equation*}
\pmb{\mathscr S}(\tau,\mathbf I(\tau))\approx\frac{(\tau-\tau_{k+1})\pmb{\mathscr S}_k-(\tau-\tau_k)\pmb{\mathscr S}_{k+1}}{\Delta\tau_k}\,,
\label{linear_interpol}
\end{equation*}
for $\tau$ in the interval $\left[\tau_k,\tau_{k+1}\right]$. One proceeds with an analytical integration and after some algebra obtains
\begin{equation}
\mathbf{\Phi}_{k+1}\mathbf I_{k+1}=\mathbf{\Phi}_k\mathbf I_k+\mathbf{\Psi}_{k+1}+\mathbf{\Psi}_k\,,
\label{delo_linear}
\end{equation}
which is an implicit linear system formally identical to Equations~\eqref{trapezoidal} and~\eqref{delo_constant}. The numerical scheme described by Equation~\eqref{delo_linear} was presented by \cite{rees+al1989} under the name of DELO and subsequently re-baptized by \cite{delacruz_rodriguez+piskunov2013} as DELO-linear. It is a second-order accurate method, as shown in Figure \ref{fig:delo1}. The explicit values of the coefficients $\mathbf{\Phi}_k$, $\mathbf{\Phi}_{k+1}$, $\mathbf{\Psi}_k$, and $\mathbf{\Psi}_{k+1}$ are provided in Appendix~\ref{appendix:B}.

The natural successive step is a parabolic Lagrangian interpolation of the effective source function \citep{murphy1990}, i.e., Equation~\eqref{lagrange} with $q=2$. Considering three spatial points $\{\tau_{k-1},\tau_k, \tau_{k+1}\}$ located along the optical depth grid, the effective source function $\pmb{\mathscr S}$ is approximated by the parabolic interpolation,
\begin{equation*}
\begin{split}
\pmb{\mathscr S}(\tau,\mathbf I(\tau))&\approx\;\pmb{\mathscr S}_{k+1}\frac{(\tau-\tau_k)(\tau-\tau_{k-1})}{\Delta\tau_k(\Delta\tau_k+\Delta\tau_{k-1})}\\
&-\;\;\pmb{\mathscr S}_k\frac{(\tau-\tau_{k+1})(\tau-\tau_{k-1})}{\Delta\tau_k\Delta\tau_{k-1}}\\
&+\;\;\pmb{\mathscr S}_{k-1}\frac{(\tau-\tau_{k+1})(\tau-\tau_k)}{\Delta\tau_{k-1}(\Delta\tau_k+\Delta\tau_{k-1})}\,,
\label{S_parabolic_interpol}
\end{split}
\end{equation*}
for $\tau$ in the interval $\left[\tau_k,\tau_{k+1}\right]$. The necessity of a third interpolation point cannot be satisfied by numerical values at $\tau_{k+2}$, because no information is available for $\mathbf I_{k+2}$. After two integrations by parts of Equation~\eqref{delo_numerical} and some algebra, one obtains
\begin{equation}
\mathbf{\Phi}_{k+1}\mathbf I_{k+1}=\mathbf{\Phi}_k\mathbf I_k+\mathbf{\Phi}_{k-1}\mathbf I_{k-1}
+\mathbf{\Psi}_{k+1}+\mathbf{\Psi}_k+\mathbf{\Psi}_{k-1}\,.
\label{delo_parabolic}
\end{equation}
The method described by Equation~\eqref{delo_parabolic}, which might be called DELO-parabolic, is third-order accurate, as shown in Figure \ref{fig:delo1}. The coefficients $\mathbf{\Phi}_{k-1}$, $\mathbf{\Phi}_k$, $\mathbf{\Phi}_{k+1}$, $\mathbf{\Psi}_{k-1}$, $\mathbf{\Psi}_k$, and $\mathbf{\Psi}_{k+1}$ are provided in Appendix~\ref{appendix:B}. The higher order of accuracy is essential to detect, for instance, the second-order behavior of the emission vector often present in realistic atmospheric models, avoiding its possible systematic overestimation.

DELO-constant and DELO-linear formal solvers compute the Stokes $\mathbf I_{k+1}$ solely on the basis of information about the preceding Stokes value $\mathbf I_k$ and are classified as one-step methods. In this sense, one-step methods have no memory, i.e., they forget all of the prior information that has been gained. In contrast, DELO-parabolic takes into account the two most recently found Stokes vectors $\mathbf I_k$ and $\mathbf I_{k-1}$, entering in the class of the multistep methods.

This family of formal solvers can be further expanded by just increasing the interpolation degree $q$ of the effective source function. For instance, a third-order Lagrangian polynomial would generate DELO-cubic. However, the complexity of the numerical methods would increase as well and the adaptation to non-uniform grids would become gradually more cumbersome.

One should emphasize that the detailed behavior of the error curves plotted in Figures~\ref{fig:delo1} and~\ref{fig:delo2} depends on the considered atmospheric model. Indeed, the main scope of these figures is to highlight the overall order of accuracy of the various methods. It would certainly be wrong to try to reach conclusions on the performance of different methods on the basis of a qualitative comparison of small details of the error curves, obtained considering a single model atmosphere. The atmospheric model considered in this work is described in Appendix C.
\subsection{DELO-B{\'e}zier methods}\label{subsec:3.2}
The choice of the Lagrangian form for the polynomial $\mathbf P_q$ in Equation~\eqref{polynomial_approx} is certainly not univocal and the literature provides different interpolation methods. \citet{mihalas+auer1978} used the Hermitian interpolation for the integration of the scalar radiative transfer equation. An interesting set of suitable interpolants was proposed by \citet{auer2003} for the same problem: among them the monotonic Hermite interpolants recently used by \citet{ibgui2013} in the IRIS code. In the same year, \citet{delacruz_rodriguez+piskunov2013} applied B{\'e}zier polynomials to the DELO strategy, generating the quadratic and cubic DELO-B{\'e}zier methods.

A detailed description of these methods, and a comparison with other methods, such as DELO-linear and DELOPAR, can be found in the above-mentioned publications, and will not be repeated here. As shown in Figure~\ref{fig:delo2}, both quadratic and cubic DELO-B{\'e}zier methods show fourth-order accuracy. It is interesting to observe that when treating smooth functions, the B{\'e}zier curves introduced by \citet{delacruz_rodriguez+piskunov2013} are forced to be identical to the Hermite polynomials of corresponding degree by adopting very specific control points. A detailed discussion of Hermitian methods will be presented in the second paper of this series, which will focus on high-order methods.
\subsection{Particular cases: DELOPAR and BESSER}\label{subsec:3.3}
\begin{figure*}
\centering
\includegraphics[width=1.\textwidth]{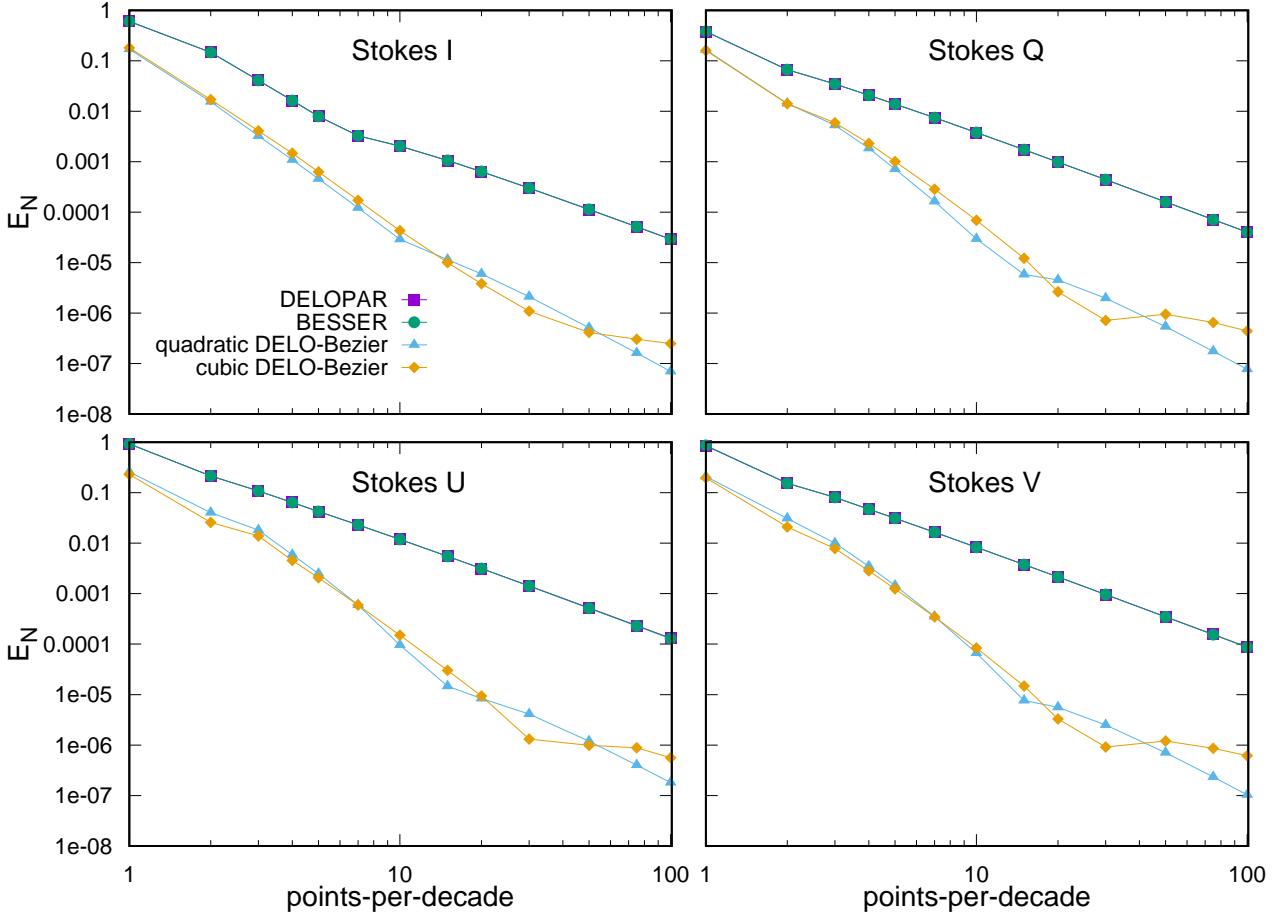}
  \caption{The log-log representation of the global error for the Stokes vector components $I,Q,U$ and $V$, as a function of the number of points-per-decade of the continuum optical depth for the DELOPAR, BESSER, quadratic, and cubic DELO-B{\'e}zier methods. The considered atmospheric model is described in Appendix~\ref{appendix:D} and the global error is computed as shown in Appendix~\ref{appendix:A}. The asymptotic behavior of the quadratic and cubic DELO-B{\'e}zier methods can only be appreciated in the interval between 3 and 11 points-per-decade.}
\label{fig:delo2}
\end{figure*}

Aiming to increase the order of accuracy with respect to the DELO-linear strategy, \cite{trujillo_bueno2003} opted for a \emph{semi-parabolic} interpolation, namely a parabolic interpolation for the modified emission vector $\tilde{\boldsymbol{\epsilon}}$ and a linear interpolation for the $\pmb{\mathscr K}\mathbf I$ term. The parabolic interpolation of $\tilde{\boldsymbol{\epsilon}}$ is performed by considering the three spatial points $\{\tau_k,\tau_{k+1}, \tau_{k+2}\}$, which differs from the set used by DELO-parabolic. In the case of a diagonal dominant propagation matrix, this strategy can increase the order of accuracy in the Stokes parameter $I$, provided a high-order integration of the opacity (see Appendix~\ref{appendix:C}). However, the local truncation error in the Stokes $Q$, $U$, and $V$, is still dominated by the linear approximation of the term $\pmb{\mathscr K}\mathbf I$, resulting in a second-order method as shown in Figure \ref{fig:delo2}. Therefore, the lack of improvement with respect to DELO-linear comes directly from the design of the method and not, as erroneously conjectured, from the so-called overshooting. This should also explain the unsatisfactory performance of DELOPAR found by \cite{delacruz_rodriguez+piskunov2013}, as presented in their error curves. The DELOPAR strategy remains a honest method for the non-polarized case or for vanishing dichroism and anomalous dispersion coefficients. In fact, in both cases, the modified propagation matrix $\pmb{\mathscr K}$ disappears and a parabolic approximation of the source term $\tilde{\boldsymbol{\epsilon}}$, supported by a proper conversion to optical depth (see Appendix~\ref{appendix:C}), produces an effective third-order numerical scheme. This can be appreciated in \cite{stepan+trujillo_bueno2013}, where the log-log error figure clearly shows the superiority of DELOPAR over DELO-linear for the scalar case. \cite{stepan+trujillo_bueno2013} applied a similar technique to create BESSER, the formal solver used in the PORTA code. The same argumentation can be applied to it, explaining its second-order accuracy confirmed by Figure \ref{fig:delo2}.
\begin{table}
\caption{Order of accuracy for DELO methods}
\setlength{\tabcolsep}{5pt}\renewcommand{\arraystretch}{1.5}
\centering
\begin{tabular}{|l|c|c|}\hline%\toprule
\emph{Formal solver}	& \emph{Order of accuracy}\\
\hline %\midrule
DELO-constant	& 2 \\
DELO-linear	& 2 \\
DELO-parabolic	& 3 \\
DELOPAR		& 2 \\
BESSER		& 2 \\
Quadratic DELO-B\'ezier	& 4 \\
Cubic DELO-B\'ezier	& 4 \\\hline %\bottomrule
\end{tabular}
\label{tab:convergence}
\end{table}
%
% %%%%%%%%%%%%%%%%%%%%%%%%%%%%%%%%%%%%%%%%
\section{Conclusions}\label{sec:sec5}
% %%%%%%%%%%%%%%%%%%%%%%%%%%%%%%%%%%%%%%%%
This paper pays particular attention to the characterization of the different formal solvers for polarized radiative transfer. The paradigmatic analysis for numerical schemes proposed here is based on three different criteria: \emph{order of accuracy}, \emph{stability}, and \emph{computational cost}.

The order of accuracy of a numerical scheme indicates how fast the error decreases, when reducing the step size. Therefore, it can be only appreciated by considering the global (or local) error dependence on the step size and not through single Stokes profiles.

When discussing numerical methods, the term stability relates to the attenuation of numerical errors with the recursive application of a scheme and not to the erratic behavior of high-order polynomial approximation (the so-called overshooting). One must always guarantee that any possible instability is avoided, when judging the order of accuracy of a method.

The computational cost is related to the complexity of the algorithm. Therefore, one suggests to maintain as bony a method as possible, avoiding unnecessary superstructures.

In this perspective, some considerations about the DELO family can be exposed. Regarding the order of accuracy, one realizes that DELO methods do not converge better than more conventional methods: DELO-constant, DELO-linear, and DELOPAR are only second-order accurate, the two-step DELO-parabolic method effectively reaches third-order accuracy, and quadratic and cubic DELO-B{\'e}zier methods usually perform as fourth-order accurate methods, as summarized in Table \ref{tab:convergence}. One must point out that second-order accuracy is already guaranteed by the simple trapezoidal method. The discussion about stability is somehow more involved. The DELO approach is thought to remove stiffness from the problem, but no stability improvement is necessary for formal solvers having an order of accuracy $p\le2$, because the trapezoidal method already guarantees A-stability as shown in Figure~\ref{stability_trapezoidal}a. However, one could still promote the DELO strategy for high-order methods ($p\ge3$). Concerning the computational cost, DELO methods differ from standard implicit numerical methods for two reasons: the DELO strategy requires an additional conversion to optical depth (see Appendix~\ref{appendix:C}), which, however, is anyway important to mitigate the variation of the eigenvalues of the propagation operator $-\mathbf K$ along the ray path, enforcing stability. Moreover, DELO coefficients require the evaluation of exponential terms (see Appendix \ref{appendix:B}) and this results in extra computational cost, as explained in Section~\ref{subsec:1.4}.

In conclusion, the necessity of the DELO strategy for the numerical treatment of the polarized radiative transfer is not warranted for low-order methods and not well motivated for high-order methods. A second paper, focused on high-order formal solvers, will try to build a clear hierarchy with respect to order of accuracy, stability and computational cost. The effective performances when dealing with realistic atmospheric models remain to be explored.
\acknowledgments
The financial support by the Swiss National Science Foundation (SNSF) through grant ID 200021\_159206/1 is gratefully acknowledged. Special thanks are extended to F. Calvo, A. Paganini, and F. Z\"uger for particularly enriching discussions. The authors are grateful to the anonymous referee for providing valuable comments that helped improving the article.
\appendix
\section{Conversion to optical depth}\label{appendix:C}
As already pointed out by \citet{delacruz_rodriguez+piskunov2013}, many radiative transfer applications require a conversion of the spatial scale, e.g., from geometrical height $s$ to optical depth $\tau$. From Equation~\eqref{opt_depth} one obtains
\begin{equation}
\Delta \tau_k = \tau_k -\tau_{k+1} = \int_{s_k}^{s_{k+1}}\eta_I(s){\rm d}s\,.
\label{conversion_opt_depth}
\end{equation}
The numerical integration introduces an error, which could lead to a reduced order of accuracy of the formal solver. In practice, a trapezoidal integration of Equation~\eqref{conversion_opt_depth} is inadequate to perform numerical schemes based on high-order interpolations of the effective source function (e.g., DELO-parabolic). Therefore, high-order DELO methods require a corresponding high-order numerical evaluation of the integral in Equation~\eqref{conversion_opt_depth}.
\section{DELO coefficients}\label{appendix:B}
In order to keep the notation as close as possible to \citet{rees+al1989}, the following definitions are introduced
\begin{align*}
E_k&=e^{-\Delta\tau_k}\,,\\
F_k&=1-E_k\,,\\
G_k&=[1-(1+\Delta\tau_k)E_k]/\Delta\tau_k\,.
\end{align*}
The coefficients of the DELO-constant method, Equation~\eqref{delo_constant}, are given by
\begin{align*}
\mathbf{\Phi}_k&=E_k\mathbf{1}-\frac{F_k}{2}\pmb{\mathscr K}_k\,,\\
\mathbf{\Phi}_{k+1}&=\mathbf{1}+\frac{F_k}{2}\pmb{\mathscr K}_{k+1}\,,\\
\mathbf{\Psi}_k&=\frac{F_k}{2}\tilde{\boldsymbol{\epsilon}}_k\,,\\
\mathbf{\Psi}_{k+1}&=\frac{F_k}{2}\tilde{\boldsymbol{\epsilon}}_{k+1}\,.
\end{align*}
The coefficients of the DELO-linear method, Equation~\eqref{delo_linear}, are given by
\begin{align*}
\mathbf{\Phi}_k&=E_k\mathbf{1}-G_k\pmb{\mathscr K}_k\,,\\
\mathbf{\Phi}_{k+1}&=\mathbf{1}+(F_k-G_k)\pmb{\mathscr K}_{k+1}\,,\\
\mathbf{\Psi}_k&=G_k\tilde{\boldsymbol{\epsilon}}_k\,,\\
\mathbf{\Psi}_{k+1}&=(F_k-G_k)\tilde{\boldsymbol{\epsilon}}_{k+1}\,.
\end{align*}
The coefficients of the DELO-parabolic method, Equation~\eqref{delo_parabolic}, are given by
\begin{align*}
\mathbf{\Phi}_{k-1}&=-\Phi_{k-1}\pmb{\mathscr K}_{k-1}\,,\\
\mathbf{\Phi}_k&=E_k\mathbf{1}-\Phi_k\pmb{\mathscr K}_k\,,\\
\mathbf{\Phi}_{k+1}&=\mathbf{1}+\Phi_{k+1}\pmb{\mathscr K}_{k+1},\\
\mathbf{\Psi}_{k-1}&=\Phi_{k-1}\tilde{\boldsymbol{\epsilon}}_{k-1}\,,\\
\mathbf{\Psi}_k&=\Phi_k\tilde{\boldsymbol{\epsilon}}_k\,,\\
\mathbf{\Psi}_{k+1}&=\Phi_{k+1}\tilde{\boldsymbol{\epsilon}}_{k+1}\,,
\end{align*}
with
\begin{align*}
\Phi_{k-1}&=\frac{-\Delta\tau_k(1+E_k)+2F_k}{\Delta\tau_{k-1}(\Delta\tau_{k-1}+\Delta\tau_k)}\,,\\
\Phi_k&=-E_k+\frac{\Delta\tau_{k-1}+\Delta\tau_k-E_k(\Delta\tau_{k-1}-\Delta\tau_k)-2F_k}{\Delta\tau_{k-1}\Delta\tau_k}\,,\\
\Phi_{k+1}&=1+\frac{E_k\Delta\tau_{k-1}-(2\Delta\tau_k+\Delta\tau_{k-1})+2F_k}{\Delta\tau_k(\Delta\tau_{k-1}+\Delta\tau_k)}.
\end{align*}
The DELO-parabolic coefficients $\Phi_{k-1},\Phi_k,\Phi_{k+1}$ could suffer of problematic division with vanishingly small quantities. Thus, in case of small $\Delta \tau$, a Taylor expansion of the exponential term $E_k$ to third-order is indicated.
\section{Atmospheric model}\label{appendix:D}
The atmosphere model and parametric description used for the calculations shown in this paper are very similar to the ones used by \citet{steiner2016}. The radiative transfer is computed at different frequencies $\nu$ in a spectral interval containing a hypothetical, magnetically sensitive spectral line. The problem is formulated in reduced frequencies
\begin{equation}
v = \frac{\nu_0-\nu}{\Delta \nu_D}\,,
\label{reduced_frequency}
\end{equation}
with $\nu_0$ the line-center frequency and $\Delta \nu_D$ the Doppler width. The spectral interval $v\in[-6,6]$ is considered. The atmosphere is assumed to be plane-parallel, and the chosen reference spatial coordinate is the continuum optical depth along the vertical direction, defined by 
\begin{equation*}
{\rm d}\tau_c=-k_c(z){\rm d}z\,,
\end{equation*}
where $k_c$ is the continuum absorption coefficient at the line-center frequency, and $z$ is the geometrical height (increasing in the outward direction). The atmosphere extends in the range $\log\tau_c\in[-5,2]$.

\begin{figure}
\vspace{0.3cm}
\plotone{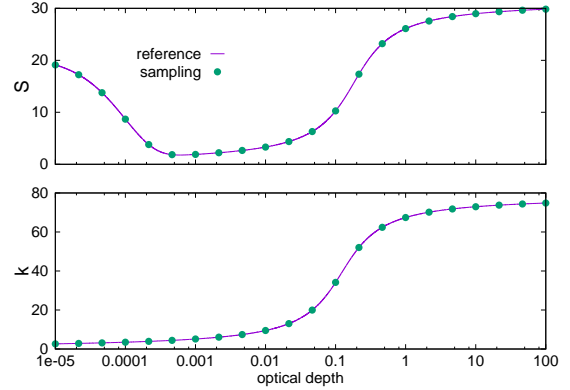}
  \caption{The source function $S$ and the ratio $k$ according to Equations \eqref{smooth1}-\eqref{smooth2}. The green dots represent a sampling with 3 points-per-decade.}
\label{fig:profile}
\end{figure}

No scattering or atomic polarization is taken into account, so that polarization is only introduced by the Zeeman effect. The magnetic field vector and the absorption and anomalous dispersion profiles (which enter in the definition of the coefficients of the propagation matrix $\mathbf K$ and of the emission vector $\boldsymbol \epsilon$) are assumed to be depth-independent. Under these assumptions, the propagation matrix can be parametrized in the form \citep[see, e.g.,][]{landi_deglinnocenti+landolfi2004,steiner2016}
\begin{equation*}
\mathbf K(\tau_c)=\mathbf 1 + k(\tau_c)\mathbf H\,,
\end{equation*}
where $k$ is the ratio between the frequency-integrated line absorption coefficient and the continuum absorption coefficient, and $\mathbf H$ is a constant $4 \times 4$ matrix. The emission vector, on the other hand, can be written as
\begin{equation*}
\boldsymbol{\epsilon}(\tau_c)= \mathbf K(\tau_c)\mathbf S(\tau_c)\,,\text{ with } \mathbf S(\tau_c)=(S(\tau_c),0,0,0)^{T}\,.
\end{equation*}
where $S$ is the usual intensity source function. The following analytical form of $S$ and $k$ (the only depth-dependent quantities of the problem) have been considered
\begin{align}
S(\tau_c)&= A_1 e^{-\tau_c/\tau_1}+A_2\cdot\left(\pi/2+\arctan(\log\frac{\tau_c}{\tau_2})\right)\,,\label{smooth1}\\
k(\tau_c)&= B\cdot\left(\pi/2+\arctan(\log\frac{\tau_c}{\tau_k})\right)\,.\label{smooth2}
\end{align}
The exponential term in Equation~\eqref{smooth1} is included in order to reproduce a possible emission rise at low optical depths (e.g., at chromospheric heights). The results of Figures~\ref{fig:delo1} and~\ref{fig:delo2} have been obtained on the basis of the smooth variation of $S$ and $k$ shown in Figure~\ref{fig:profile}, with the following values of the parameters appearing in Equations~\eqref{smooth1} and~\eqref{smooth2}: $A_1= 20$, $A_2=10$, $\tau_1=10^{-4}$, $\tau_2=0.183$, $B =25$, and $\tau_k=0.123$.

The intensity of the magnetic field is specified through the dimensionless parameter $v_B = \nu_L / \Delta \nu_D$, $\nu_L$ being the Larmor frequency. The results shown in Figures~\ref{fig:delo1} and~\ref{fig:delo2} have been obtained for $v_B = 1.5$, and assuming the magnetic field to have an inclination $\theta = 60^{\circ}$ with respect to the vertical. A damping constant $a=0.05$ has been chosen for calculating the Voigt and Faraday-Voigt profiles entering the definition of the radiative transfer coefficients. No macroscopic (bulk) velocity has been considered. The calculations have been performed for the radiation propagating outwards  in the atmosphere, along the vertical direction. At the bottom of the atmosphere (boundary condition) an unpolarized radiation field $\mathbf{I}_0 = (B_0, 0, 0, 0)^T$ has been introduced. The Stokes parameters of the emergent radiation have been calculated in units  of the parameter $B_0$, whose exact value is thus irrelevant for the calculations shown in this paper. The reference direction for positive Stokes $Q$ has been taken in the plane defined by the vertical and by the magnetic field.
\section{Error calculation}\label{appendix:A}
Denoting with $\mathbf I^{\rm ref}(v)$ and $\mathbf I^{\rm num}(v)$ the reference and the numerically computed emergent Stokes vectors, respectively, at the reduced frequency $v$, the global error for the $i$th Stokes vector component is computed as 
\begin{equation}
E_i=\frac{\displaystyle \max_{v}| I_i^{\rm ref}(v)- I_i^{\rm num}(v)|}{{\displaystyle \max_{v}} \; I_i^{\rm ref}(v)-{\displaystyle\min_{v}} \; I_i^{\rm ref}(v)}\,.
\label{error}
\end{equation}
The error is given by the maximal discrepancy between the reference and the simulated Stokes parameter over the entire profile, normalized by the maximal amplitude in the reference profile. Equation~\eqref{error} is not defined for a constant profile, because of a vanishing denominator. In that case, one needs to introduce a different error definition. The reference emergent Stokes profile $\mathbf I^{\rm ref}(v)$ is the exact solution approximated by means of high-order numerical methods, using a hyperfine grid sampling with more than $10^3$ points-per-decade of the continuum optical depth. Different high-order methods (e.g., DELO-parabolic and quadratic DELO-B{\'e}zier) are used to cross-check the reference emergent profile.
%
%% The following command ends your manuscript. LaTeX will ignore any text
%% that appears after it.
\bibliographystyle{apj} 
\bibliography{bibfile}
\end{document}